\documentclass{PoS}
\usepackage{graphicx}
\newcommand{\be}{\begin{equation}}
\newcommand{\ee}{\end{equation}}
\newcommand{\bea}{\begin{eqnarray}}
\newcommand{\eea}{\end{eqnarray}}
\newcommand{\F}{\phantom {1}}

\PoS{PoS(LAT2005)300}

\title{%
\begin{picture}(0,0)(0,0)%
\put(0,75){\makebox(0,0)[l]{\textnormal{\normalsize DESY 05-180}}}%
\put(0,60){\makebox(0,0)[l]{\textnormal{\normalsize HU-EP-05/54}}}%
\end{picture}%
Localization properties of the
topological charge density and the low lying eigenmodes of
overlap fermions}

\ShortTitle{Localization properties of the
topological charge density\ldots
}

\author{\speaker{Yoshiaki Koma}$^{a}$,
Ernst-Michael~Ilgenfritz$^{b}$, Karl~Koller$^{c}$, Gerrit~Schierholz$^{a,d}$,
Thomas~Streuer $^{d,e}$,  Volker~Weinberg$^{d,e}$\\
~\\
$^{a}$ Deutsches Elektronen-Synchrotron DESY, 22603 Hamburg, Germany\\
$^{b}$ Institut f\"ur Physik, Humboldt Universit\"at zu Berlin,
12489 Berlin, Germany\\
$^{c}$ Sektion Physik, Universit\"at M\"unchen,
80333 M\"unchen, Germany\\
$^{d}$ John von Neumann-Institut f\"ur Computing NIC / DESY,
15738 Zeuthen, Germany\\
$^{e}$ Institut f\"ur Theoretische Physik, Freie
Universit\"at Berlin, 14196 Berlin, Germany}

\author{QCDSF Collaboration\\
~\\
}

\abstract{
Overlap fermions, which preserve exact chiral symmetry
on the lattice, provide a powerful tool for investigating the
topological structure of the vacuum.
Applying this formulation to zero-temperature quenched SU(3)
configurations generated by means of the L\"uscher-Weisz action,
we define the topological charge density with and without
UV filtering and study its properties by looking at the
density profile and the two-point correlation function.
We observe that the density possesses
global sign coherent structures,
which get increasingly tangled as more and more modes are included.
This change of the structure is also detected by the increasing
negative tail of the two-point function.
We also study the inverse participation ratio of the eigenmodes
and discuss their dimensionality.
}

\FullConference{XXIIIrd International Symposium
on Lattice Field Theory\\
25-30 July 2005\\
Trinity College, Dublin, Ireland}

\begin{document}

\section{Introduction}

\par
Topological excitations,  phenomenologically
modeled as an interacting ensemble of instantons and anti-instantons,
are believed to play a prominent role in the low-energy behavior of QCD,
especially in realizing the axial U(1) anomaly (large $\eta'$ mass) and
chiral symmetry breaking.
It is thus of great interest to understand the topological structure of
the QCD vacuum from first principles and to clarify whether instantons
are indeed providing a realistic description.

\par
For this purpose, we employ the overlap fermion Dirac (Neuberger-Dirac)
operator defined by
\bea
D =\frac{\rho}{a}(1+\frac{X}{\sqrt{X^{\dagger}X}}) \; ,
\quad
X=D_{W}-\frac{\rho}{a}\; ,
\eea
where $D_{W}$ is the Wilson-Dirac operator ($r=1.0$).
We set $\rho=1.4$ as the optimal choice.
Overlap fermions possess exact chiral symmetry on the lattice
and provide $n_{-} + n_{+}$ exact zero modes, $D \psi_{n}^{\pm}=0$,
with $n_{-}$ ($n_{+}$) being the number of modes
with negative (positive) chirality
$\gamma_{5} \psi_{n}^{-}= -\psi_{n}^{-}$
($\gamma_{5} \psi_{n}^{+}= +\psi_{n}^{+}$).
The index is given by $Q = n_{-} - n_{+}$.
The non-zero modes with eigenvalue $\lambda$,
$D \psi_{\lambda} = \lambda \psi_{\lambda}$,
occur in complex conjugate pairs
$\lambda$ and $\lambda^{*}$ and satisfy
$(\psi_{\lambda}^{\dagger}, \gamma_{5}\psi_{\lambda})=0$.
The topological charge density $q(x)$, which satisfies the
index theorem with $Q= \sum_{x} q(x)$, is  defined by
\bea
q(x) \equiv
- \mbox{Tr} \left [ \gamma_{5}  (1  -\frac{a}{2} D(x,x) )\right ] \; ,
\label{eqn:q}
\eea
where `Tr' is taken for color and spinor indices.
The role of the low-lying Dirac eigenmodes for $q(x)$
is exposed by applying the eigenmode expansion,
where the cutoff $\lambda_{\rm cut}$ implies
a kind of UV filtering,
\bea
q_{\lambda_{\rm cut}}(x) = -
\sum_{|\lambda|\leq|\lambda_{\rm cut}|}
\left( 1-\frac{\lambda}{2} \right )
 c_{\lambda}(x) \; ,
\label{eqn:q-expansion}
\eea
where $c_{\lambda}(x)= \psi_{\lambda}^{\dagger}(x)
\gamma_{5}\psi_{\lambda}(x)$
is the local chirality of the mode with eigenvalue $\lambda$.
Note that the UV filtering maintains the index theorem
independently of the cutoff
such that $Q=\sum_{x}q_{\lambda_{\rm cut}}(x)$,
while controlling the UV fluctuation of the density.
This is because the index is computed only from
the zero modes which actually occur 
all with the same chirality.

\section{Observables and results}

\par
In this presentation we study the topological charge density
as well as its two-point correlation function.
Furthermore we investigate
the inverse participation ratio (IPR) of the eigenmodes.
We use an ensemble of zero-temperature
quenched configurations generated
by means of the L\"uscher-Weisz gauge action (see
Table~\ref{tab:configuration} and ref.~\cite{Galletly:2003vf}).
This action is suitable for topological studies since
dislocations are greatly suppressed.

\begin{table}[hbt]
    \centering
    \caption{Simulation details: the L\"uscher-Weisz gauge action
   is used}
\begin{tabular}{|c|c|cc|c|c|}
\hline
$\beta$ & $a$ [fm]  & $V=L^{3}T$  & (fm$^{4}) $
& \# of conf. & \# of eigenmodes\\
\hline
8.45 & 0.095  & $12^{3}24$&(3.38) &  116 &$O(50)\F$  \\
\hline
8.45 & 0.095  &  $16^{3}32$  & (10.6)  &267  & $O(140)$ \\
\hline
8.45 & 0.095  &  $24^{3}48$ & (54.0)  &  186  &$O(160)$  \\
\hline
8.10 & 0.125 & $12^{3}24$  &(10.1)  &  254 & $O(140)$ \\
\hline
\end{tabular}
\label{tab:configuration}
\end{table}

\begin{figure}[t]
    \centering
\includegraphics{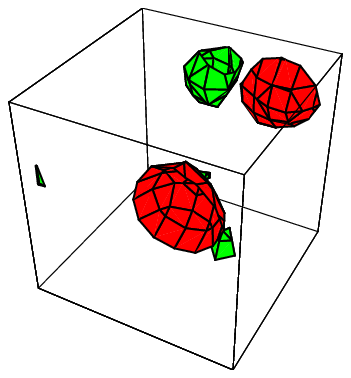}
\includegraphics{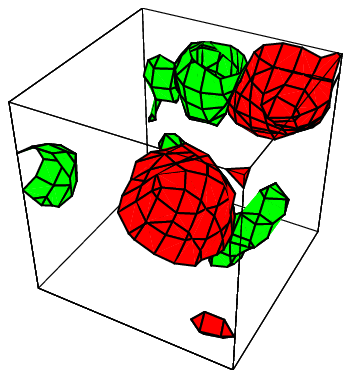}
\includegraphics{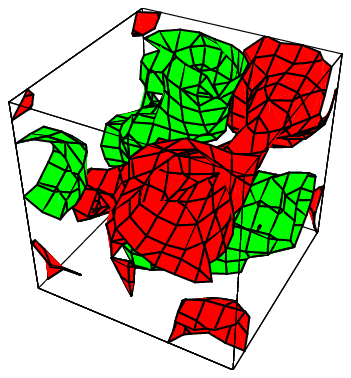}
\includegraphics{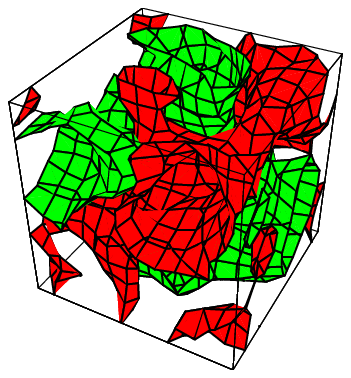}
\caption{Distribution of the topological
charge density in a given time slice of a $12^{3}24$ lattice at
$\beta=8.10$:
cutoff Im~$\lambda_{\rm cut}=$0.14, 0.28, 0.42, 0.56
(from left to right), isosurfaces at $q(x)=\pm 0.0005$ (red/green).}
\label{fig:fig1}
\end{figure}

\begin{figure}[t]
    \centering
\includegraphics{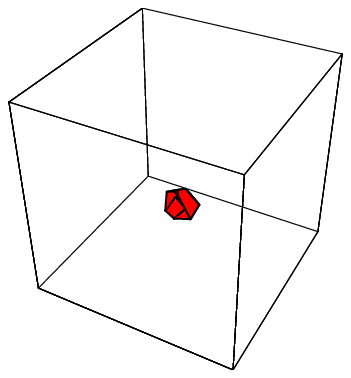}
\hspace*{1cm}
\includegraphics{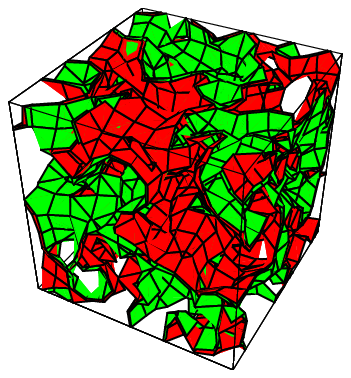}
\caption{The same plot as in Fig.~1
for the zero mode contribution (left) and the full density (right).}
\label{fig:fig2}
\end{figure}

Let us first look at the topological charge density.
In Fig.~\ref{fig:fig1}, we show the density
in a given time slice on a $12^{3}24$ lattice
at $\beta=8.10$.~\footnote{3D movies available on request.}
The red~(green) color surfaces are
the isosurfaces corresponding to $q(x)=\pm 0.0005$.
Periodic boundary conditions are imposed.
In Fig.~\ref{fig:fig2}, we show the same plot for the zero-mode
contribution alone and for the full density (no cutoff).
It is apparent that the topological charge density
possesses global sign-coherent structures (clusters).
The sizes of the clusters increase as the cutoff $\lambda_{\rm cut}$
is relaxed.
This feature is common also to the other configurations.
We will address the local tomography of the clusters
of topological charge in a future publication.

\begin{figure}[t]
\centering
\includegraphics[width=7.5cm]{lat05_qq_b810_1224.EPSF}
\includegraphics[width=7.5cm]{lat05_qq_b810_1224_focus.EPSF}
\caption{Two-point function of the topological charge density
$q_{\lambda_{\rm cut}}$
on the $12^{3}24$ lattice at $\beta=8.10$ defined 
for various $\lambda_{\rm cut}$, compared to that of the full
density $q$ ($N_{\rm conf}=3$).
The two-point function is averaged over pairs $(x,y)$ with 
the same distance $R$.
The right figure focusses at the region in $R$ where the cutoff-related
correlator turns negative. We obtained a similar picture at $\beta=8.45$.}
\label{fig:fig3}
\includegraphics[width=7.5cm]{lat05_qq_fullcomp.EPSF}
\includegraphics[width=7.5cm]{lat05_qq_fullcomp_focus.EPSF}
\caption{Two-point function of the full topological density
on the $12^{3}24$ lattice at $\beta=8.10$
($N_{\rm conf}=3$) and
on the $16^{3}32$ lattice at $\beta=8.45$
($N_{\rm conf}=2$).
The right figure shows the same as the left but magnified.
Since only a few configurations were available for this comparison
we did not estimate the errors.}
\label{fig:fig4}
\end{figure}

\par
Let us next discuss the
two-point function of the topological charge density,
\bea
C_{q}(R) =
\frac{1}{V} \langle  \sum_{x}\; q(x) q(y)  \rangle \; ,
\eea
where $R=|x-y|$ is the Euclidean distance ($R = r/a$).
The result is shown in Fig.~\ref{fig:fig3}.
The cutoffs chosen here are corresponding to those in Fig.~\ref{fig:fig1}.
We observe that while there is always a positive core near the origin
at any value of the cutoff,
the tail is negative for sufficiently large cutoffs.
As the cutoff increases, the peak of the positive core
grows and the negative tail also increases.
In fact, this is an expected behavior from the
density distribution for various cutoffs as seen
in Figs.~\ref{fig:fig1} and~\ref{fig:fig2}.
Since the density of both signs get increasingly tangled
when more and more modes are included, any point of positive density
comes closer to points of negative density.
It is interesting to note that the topological susceptibility
$\chi_{\rm top} =\langle Q^{2} \rangle / V$
is not affected by the cutoff.
Therefore, as the core peak increases, the contribution from
the negative region {\em must} grow to maintain the sum over
$C_{q}(R)$ with respect to~$R$.
The dependence on the lattice spacing $a$ is examined in
Fig.~\ref{fig:fig4} for the full density,
where the scale is introduced by multiplication with~$a^{-8}$.
Clearly, as $a$ decreases, the core peak (size) is
growing (shrinking), which seems to approach
an expected behavior in the continuum
theory~\cite{Seiler:2001je},
although the finite 
positive core at our highest $\beta$ stays at a 
radius $R \approx 2~a$
showing the limits of locality of the operator $q(x)$
itself (see also ref.~\cite{Horvath}).

\par
\begin{figure}[t]
\centering
\includegraphics[width=13cm]{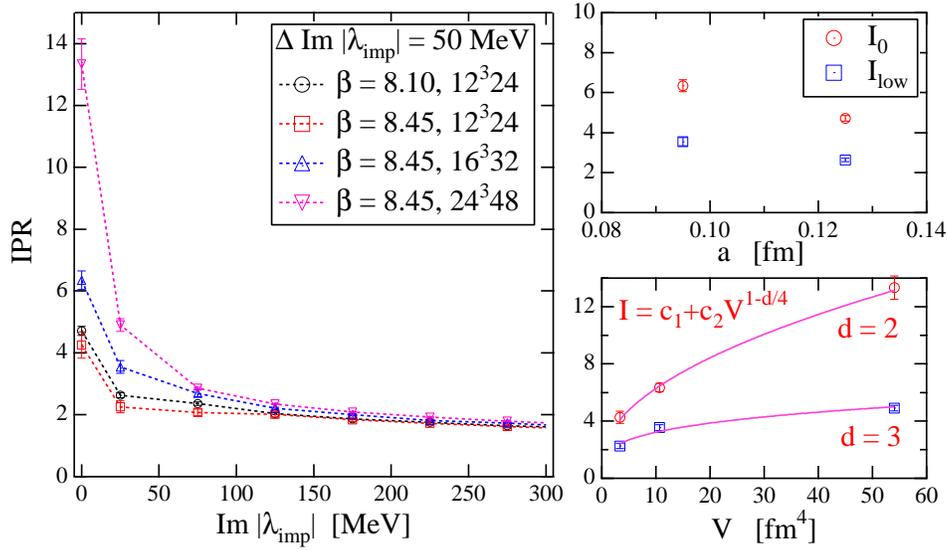}
\caption{IPR: the dependence on ${\rm Im}\lambda$ (left),
the $a$ dependence
(right upper), and the $V$ dependence (right lower).}
\label{fig:fig5}
\end{figure}

Finally let us discuss the
inverse participation ratio (IPR) of the eigenmodes.
The IPR of an eigenfunction $\psi_{\lambda}(x)$ is defined by
\bea
I = V \sum_{x \in V} \rho_{\lambda}^{2}(x) \;,
\eea
where $\sum_{x \in V}\rho_{\lambda}(x)=1$ with
$\rho_{\lambda}(x)=\psi_{\lambda}^{\dagger}(x) \psi_{\lambda}(x)$.
The IPR, $I=1/f$, denotes the inverse fraction $f$ of sites
forming the support of
$\rho_{\lambda}(x)$%
~\cite{Gattringer:2001mn,Aubin:2004mp,Gubarev:2005jm}.
Limiting cases are $I=1$ for $\rho(x)= 1/V$ (delocalized),
$I=V$ for $\rho(x)=\delta_{x,x_0}$  (extremely localized).
From the dependence on $a$ and $V$ one may infer
the effective dimension $d$ of the eigenmode
based on the relations
$I \propto a^{d-4}$ and $I \propto V^{1-d/4}$.
This means that there is no dependence on $a$ and $V$ if the mode
is extending in $d=4$ dimensions.
In Fig.~\ref{fig:fig5} (left) we show the IPR
as a function of the (imaginary part of the) eigenvalue,
where the IPR values are averaged in bins with
bin size $\Delta {\rm Im} \lambda_{\rm imp} = 50$ MeV, except
for the zero modes which are considered separately.
Here, $\lambda_{\rm imp}$ is the eigenvalue of the
improved Dirac operator $D_{\rm imp} = (1-a D /2\rho)^{-1} D$,
which projects the eigenvalue of $D$ stereographically
onto the imaginary axis.
The continuous eigenvalues of $D_{\rm imp}$ still
come in pairs $\pm i\lambda$, while the zero eigenvalues remain zero.
In Fig.~\ref{fig:fig5}, we also plot
the averaged IPR with respect to its $a$ dependence at
fixed $V$ (right upper),
and with respect to the $V$ dependence with a fixed $a$ (right lower),
separately for
zero modes ($I_{0}$) and lowest non-zero modes ($I_{\rm low}$),
the latter for $0< | {\rm Im}~\lambda_{\rm imp} | \le 50$ MeV.
These plots exhibit that the zero and lowest-lying modes
have an IPR clearly larger than one and are
sensitive to the change of $a$ and $V$, respectively.
This indicates that these modes are localized to some extent
and  possess an effective dimension less than $d=4$.
From the $V$ dependence, $I=c_{1}+c_{2}V^{1-d/4}$,
we find that $d=2$ and $d=3$
describe the IPR data well for
the zero and lowest-lying modes, respectively.
On the other hand, for the higher modes
($| {\rm Im}~\lambda_{\rm imp} | > 200$ MeV)
we find $I \lesssim 2$, practically independent on $a$ and $V$.
Hence we may conclude that these modes are delocalized
and extending in all $d=4$ dimensions.
The behavior of the IPR observed here
is qualitatively the same as found in
the SU(2) case~\cite{Gubarev:2005jm}.
However, the actual values of the IPR are
essentially smaller in our case
and our fit results are also different.

\section{Summary}

Overlap fermions preserve exact chiral symmetry on the lattice and
provide exact zero modes, which allow us to unambiguously compute the
index $Q$ of vacuum configurations. It is then possible to define the
topological charge density $q(x)$
and to expose the role of low-lying eigenmodes in $q(x)$
by applying the eigenmode expansion (UV filtering).
This approach is gauge invariant and
leaves the lattice scale unchanged in contrast to the cooling method.
In this sense, overlap fermions are a powerful tool for investigating
the topological structure of the vacuum.

\par
Using this formalism, we have investigated the localization
properties of the topological charge density by looking at the
density profile as well as the two-point correlation function
on zero-temperature quenched configurations.
We have also examined the IPR of the eigenmodes and their
dimensionality.
We have found that
the topological charge density possesses global sign coherent
structures,
which get increasingly tangled as
more and more modes are included.
Such a change of density structures has also been detected
in the behavior of the two-point function.
We note that further preliminary studies
show that the full topological charge density has a
lower-dimensional laminar structure,
together with a lumpy structure inside the sign
coherent regions.
Details will be presented in our forthcoming publication.
The IPR has indicated that the zero and low-lying modes,
typically for Im~$\lambda_{\rm imp} \lesssim 200$ MeV,
are localized to some extent and have an effective dimension
less than four.

\par
A similar investigation at finite temperature
near $T=T_{c}$ on $N_{f}=2$ dynamical gauge
configurations is discussed in ref.~\cite{Weinberg}.

\section*{Acknowledgments}

The numerical calculations have been performed at NIC (J\"ulich)
and HLRN (Berlin) as well as at DESY (Zeuthen). We thank all
institutions for support. This work has been supported in part by
the EU Integrated Infrastructure Initiative Hadron Physics (I3HP)
under contract RII3-CT-2004-506078 and by the DFG under contract
FOR 465 (Forschergruppe Gitter-Hadronen-Ph\"anomenologie).

\end{document}